\documentclass[preprint]{aastex}

\received{}
\revised{}
\accepted{}
\ccc{}
\cpright{}{}

\slugcomment{AJ, in press, May 2004}
\shorttitle{X-ray NGC2264}
\shortauthors{}

\newcommand{\etal}{{\it et al.\/}}

\newcommand{\lx}{${\rm L_x}$}
\newcommand{\lbol}{${\rm L_{bol}}$}

\begin{document}

\title{Chandra X-ray observations of Young Clusters I. NGC2264 Data.
}

\author{ 
Solange V. Ram\'{\i}rez \altaffilmark{1},
Luisa Rebull \altaffilmark{1},
John Stauffer \altaffilmark{1},
Thomas Hearty \altaffilmark{2},
Lynne Hillenbrand \altaffilmark{3},
Burton Jones \altaffilmark{4},
Russell Makidon \altaffilmark{5},
Steven Pravdo \altaffilmark{2},
Steven Strom \altaffilmark{6}, \&
Michael Werner \altaffilmark{2}
.}

\altaffiltext{1}{Spitzer Science Center, Mail Stop 220-06,
California Institute of Technology.}
\altaffiltext{2}{Jet Propulsion Laboratory}
\altaffiltext{3}{Astronomy Department, California Institute of Technology.}
\altaffiltext{4}{University of California, Santa Cruz}
\altaffiltext{5}{Space Telescope Science Institute}
\altaffiltext{6}{NOAO, Kitt Peak National Observatory}

\begin{abstract}
We present results of a Chandra observation of a field in NGC 2264.
The observations were taken with the ACIS-I camera 
with an exposure time of 48.1 ks.
We present a catalog of 263 sources, which includes X-ray luminosity,
optical and infrared photometry
and X-ray variability information. We found 41 variable sources,
14 of which have a flare-like light curve, and 2 of which
have a pattern of a steady increase or decrease over a
10 hour period.
The optical and infrared
photometry for the stars identified as X-ray sources are consistent
with most of these objects being pre-main sequence stars with ages 
younger than 3 Myr.
\end{abstract}

\keywords{stars: activity --- 
stars: pre-main sequence ---
X-rays: stars}

\section{INTRODUCTION}

NGC 2264 is a cluster of young stars, part of the Mon OB1 association.
It is about 3 Myrs old \citep{par00}, and the shape of its initial
mass function is very similar to that of Orion \citep{sun97}.
Pre-main sequence (PMS) stars have been identified in NGC 2264 
using a variety of methods, including $H\alpha$ spectroscopy 
\citep{her54,ogu84}, $H\alpha$ narrow band photometry \citep{sun97,par00},
irregular variability \citep{ada83}, near IR photometry \citep{lad93},
proper motions \citep{vas65}, and X-ray flux \citep{fla00,pat94,sim85}.
Furthermore, several techniques have been used to identify circumstellar
disk candidates, including excess ultraviolet ($U-V$) emission,
excess near IR ($I-K$ and $H-K$) emission, and $H\alpha$ emission line
equivalent widths \citep[e.g. ][]{reb02}. 
\citet{reb02} have found a good correlation between disk indicators 
and report lower limits for the disk fraction ranging from 21\% to 56\%.
They also found a typical value for mass accretion rates of $\sim 10^{-8}
{\rm M_{\odot}~yr^{-1}}$, comparable to the values derived for Orion
and Taurus-Auriga.

NGC 2264 provides a laboratory for studying the interrelationships of 
rotation, mass accretion, disk indicators and X-ray luminosity of PMS stars. 
The question of how exactly these things are related is still an open 
problem in star formation phenomenology. 
There is a clear relationship between rotation rate (period) and
X-ray luminosity (\lx) found in late-type stars in clusters as old as
NGC 2547 \citep[15-40 Myrs,][]{jef00} through the Hyades 
\citep[$\sim$500 Myrs,][]{sta97}.
The ratio between the X-ray and bolometric luminosity, \lx/\lbol, 
increases with increasing rotation rate, until the most rapidly 
rotating stars reach a maximum X-ray luminosity (or saturation level) 
such that \lx/\lbol $\sim 10^{-3}$ 
\citep[see ][and references within]{piz03}. 
It is much less clear that rotation and \lx/\lbol~ are related in
younger clusters \citep[e.g. ][]{gag95}.   
\citet{fei03} see no obvious correlation between rotation and \lx/\lbol~ 
for their Orion \citep[$\sim$ 1 Myr old; ][]{hil97} sample, and conclude 
that the X-ray generation mechanism
for young PMS stars must be different from that responsible in
young main sequence stars.   
\citet{fla03} analyze data for a number of young associations and 
clusters (including Orion), and agree that there is little correlation 
between \lx/\lbol~  and rotation at very young ages, but conclude 
that the data are consistent with a single physical mechanism, where 
the Orion-age stars are simply all at or near the saturation level 
(and that level and the critical velocity for it are a function of 
gravity/age).
We want to study at what age the relationship between rotation 
and \lx/\lbol~ emerges.
NGC 2264 \citep[$\sim$ 3 Myrs old; ][]{sun97,par00}, being slightly older 
than Orion, is at an ideal age to probe the
relationship between these parameters.

Chandra observations provide a unique tool to improve the X-ray sample 
of PMS stars in NGC 2264.
The high spatial resolution provided by the ACIS camera allows us to 
resolve the source confusion present in previous ROSAT data samples.
The Chandra sensitivity extends the X-ray flux limit from the ROSAT value
of log\lx(erg/s)$\sim$30.1  to log\lx(erg/s)$\sim$28.5,
allowing us to detect lower mass stars and explore the mass dependence of
the rotation and \lx~ relationship.

In the present paper, we present results of Chandra observations
of NGC 2264.
We discuss source detection, variability and \lx~ determination,
providing a catalog of 263 X-ray sources.
In paper II \citep{reb03}, we will discuss in more detail the 
relationships found here between rotation rate, mass accretion rate, 
disk indicators, and X-ray luminosity.  
We rely heavily on data from our earlier papers in this
cluster, \citet{reb02} and \citet{mak04}, providing optical
photometry and periods, respectively.

\section{OBSERVATIONS}

\begin{figure}
\epsscale{0.7}
\plotone{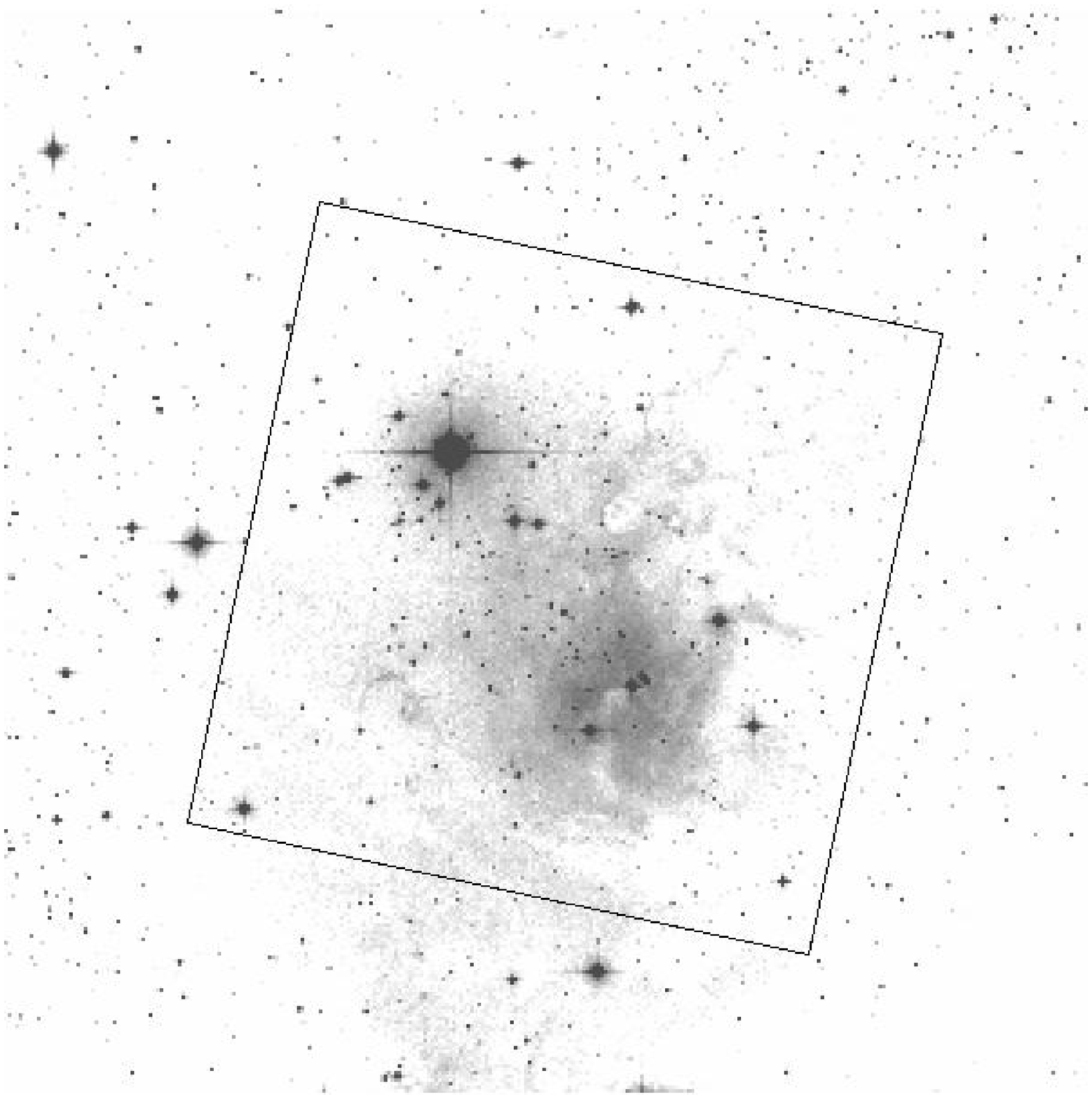}
\caption[dss_field.ps]{Image of NGC 2264 from the Palomar Digital Sky Survey
\citep{rei91}.
The image has a field of view of $30\arcmin \times 30\arcmin$ and the box 
represents the field of view of the Chandra ACIS-I observations, centered at 
RA(2000)=$6^{h}40^{m}48^{s}$, DEC(2000)=$+9\arcdeg51\arcmin$.
\label{dss_field}}
\end{figure}

NGC 2264 was observed with the Advanced CCD Imaging Spectrometer
(ACIS) detector on board the {\it Chandra X-ray Observatory} \citep{wei02}
on 2002 February 9. 
The results presented here arise from the imaging array (ACIS-I), which
consists of four 1024$\times$1024 front-side illuminated CCDs.
The array is centered at $6^{h}40^{m}48^{s}, +9\arcdeg51\arcmin$' 
and covers an area on the sky of about $17\arcmin \times 17\arcmin$. 
Figure ~\ref{dss_field} shows a $30\arcmin \times 30\arcmin$ image of 
NGC 2264 from the Palomar Digital Sky Survey \citep{rei91}, with the 
Chandra ACIS-I field of view marked as a box.
The field was selected to maximize the number of stars
in the field of view with known periods from \citet{mak04} and 
with minimal overlap with another contemporaneous Chandra NGC2264 
observation \citep{fla03}.
The total exposure time of the ACIS observations is 48.1 ks.

\subsection{Data Preparation}\label{data}

\begin{figure}
\epsscale{0.7}
\plotone{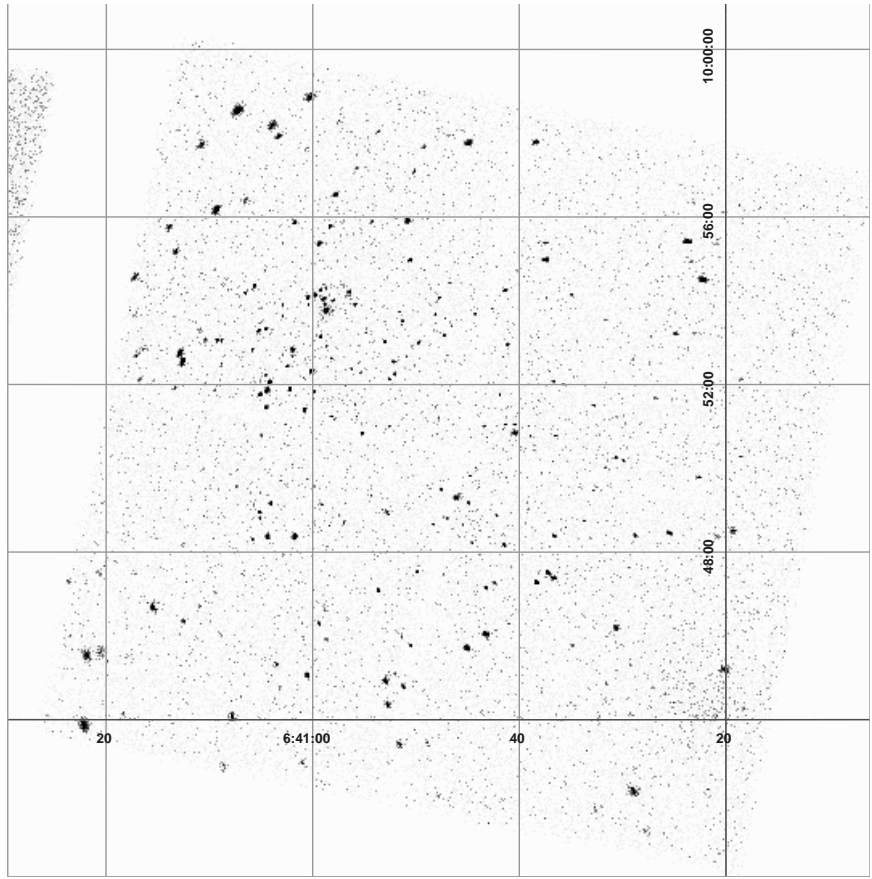}
\caption[image.ps]{Image of NGC 2264, observed with ACIS-I at the 
Chandra Observatory. The image has a field of view of 17'$\times$17' 
and it is centered at RA(2000)=$6^{h}40^{m}48^{s}$, 
DEC(2000)=$+9\arcdeg51\arcmin$.
This image contains only filtered events.
\label{image}}
\end{figure}

We started data analysis with the Level 1 processed event list provided
by the pipeline processing at the {\it Chandra} X-ray Center. 
We have kept all events, including the ones flagged as cosmic ray
afterglows by the pipeline. 
The presence of spurious sources due to cosmic ray afterglows was 
eliminated using the light curves of the detected sources 
(see Sec.~\ref{variability}).
The energy and grade of each data event were corrected for charge
transfer inefficiency (CTI), applying the algorithm code described in
\citet{tow00}.
Then, the event file was filtered by ASCA grades (keeping
grades 0, 2, 3, 4, and 6); by time intervals; and by background
flaring due to solar activity.   Removal of the background flares
identified by the latter step reduced the exposure time
to 47.4 ks.
The filtering process was done using the Chandra Interactive Analysis 
of Observations (CIAO) 
package\footnote{http://cxc.harvard.edu/ciao/index.html} 
and following the 
Science Threads\footnote{http://cxc.harvard.edu/ciao/threads/index.html} 
provided by the {\it Chandra} X-ray Center.
Finally, the energy range was restricted to 0.3 -- 10 keV.
Figure ~\ref{image} shows the ACIS-I image of the filtered observations.

\subsection{Source detection}

The CIAO package provides three source detection tools, as part of
their $Detect$ software\footnote{http://cxc.harvard.edu/ciao/download/doc/detect\_html\_manual}.
The $celldetect$ tool uses a sliding square cell to search for statistically 
significant enhancements over the background. This tool has been widely used
in Einstein and ROSAT data, and works well in the detection of faint sources
outside crowded fields.
The $vtpdetect$ tool determines individual densities for every occupied
pixel, and analyzes the distribution of densities for significant source 
enhancements. This tools works well in the detection of low surface
brightness features, but combines closely spaced point sources.
The $wavdetect$ tool  performs a Mexican hat wavelet decomposition and
reconstruction of the image and then searches for significant correlations. 
This tool works well in separating closely spaced point sources. 
We choose the $wavdetect$ tool to determine the X-ray sources in our
$Chandra$ observations, since it is the tool that properly handles crowded
fields like ours.
The algorithm used by $wavdetect$ is described in \citet{fre02}. 
We used wavelet scales ranging from 1 to 16 pixels in steps of $\sqrt{2}$,
and the default source significance threshold of $1\times10^{-6}$.
The {\it wavdetect} tool was run separately in the four ACIS-I CCDs images
and it produced an original list of 313 sources (see Sec~\ref{xray_phot}). 

\subsection{Astrometric Alignment}

\begin{figure}
\epsscale{0.7}
\plotone{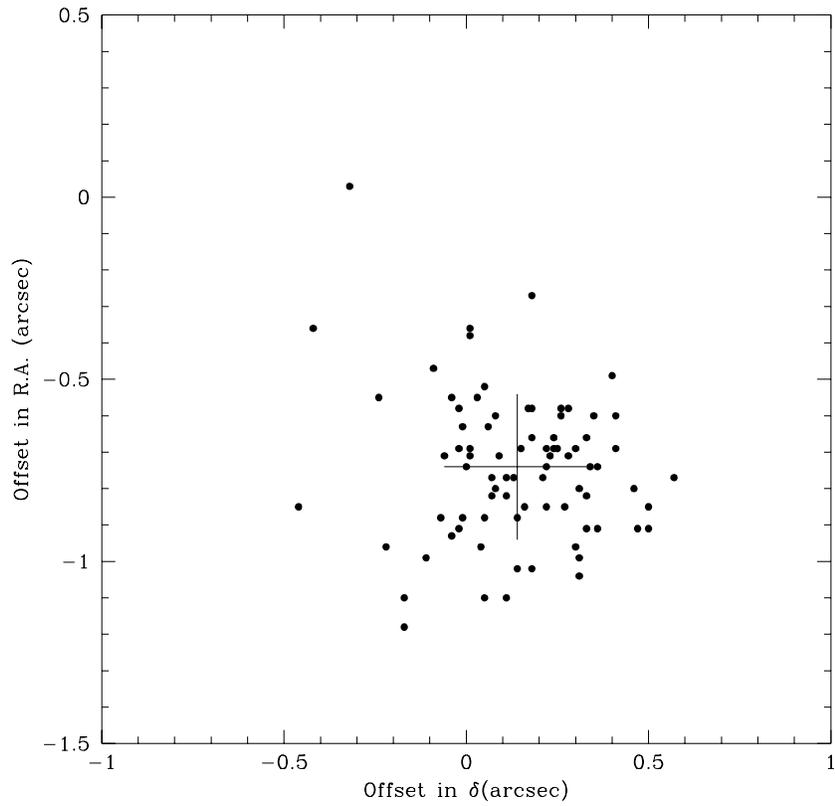}
\caption[offset.ps]{Position offsets of 80 X-ray sources, located within
$\phi < 5'$, with respect to positions from $I_c-$band observations. 
The mean offset in RA and DEC is marked by a cross, the length of which is 
equal to the standard deviation around the mean RA and DEC.
\label{offset}}
\end{figure}

The positions of the sources obtained by {\it wavdetect} were correlated
with $I_c-$band positions \citep{reb02} from the optical/infrared catalog 
of NGC 2264
stars compiled for our project (see description of the catalog in
Sec.~\ref{catalog}).
Each X-ray source was manually checked to confirm that the optical/infrared 
counterpart lay within the radius for count extraction (defined in
Sec.~\ref{xray_phot}).
We checked the astrometry of the Chandra observations using all X-ray 
sources located within an off-axis angle ($\phi$) less than 5 arcmin 
that have $I_c-$band counterparts from \citet{reb02}.
A total of 80 sources meet this criteria.
We determined a mean offset in R.A. of --0.''74 $\pm$ 0.20 and a mean 
offset in DEC of 0.''14 $\pm$ 0.20 between the Chandra and $I_c-$band
coordinates.
Figure ~\ref{offset} shows the coordinate offsets of the 80 sources 
within $\phi < 5'$ and with $I_c-$band counterparts. 
The mean offset is marked by a cross. 
The Chandra positions were corrected by these mean offsets 
to have the X-ray sources in the same reference frame as their optical 
counterparts.

\section{RESULTS}

\subsection{X-ray Photometry}\label{xray_phot}

X-Ray aperture photometry was initially performed on the 313 sources detected
by the CIAO tool {\it wavdetect}. 
The radius for count extraction, $R_{ext}$, is different for each source, 
given the variation of the point spread function (PSF) across the field. 
\citet{fei02} defined the 95\% and 99\% encircled energy radii
($R$(95\%EE) and $R$(99\%EE)),
as a function of off-axis angle ($\phi$), in their footnote number 12.
We set $R_{ext}$ equal to $R$(95\%EE), except for
sources with higher than 1000 counts, where we enlarged
$R_{ext}$ to be equal to $R$(99\%EE).
The X-ray counts ($C_{extr}$) were extracted within a circular region
of radius $R_{ext}$, using the CIAO tool {\it dmextract}.
We also defined an annulus around each source for background determination.
The annulus is defined to be between 1.2$R$(99\%EE) and 1.5$R$(99\%EE). 
The background counts were also extracted using the CIAO tool {\it dmextract}.
We computed the background in counts$\times$arcsec$^{-2}$ as a function
of off-axis angle ($\phi$), performing a 3$\sigma$ rejection fit to
avoid background counts from annuli that have other sources within.
Indeed, 38 annuli contained detected sources. 
The computed background was constant as a function of $\phi$, and 
it had a value of $B$=(0.063$\pm$0.023)counts$\times$arcsec$^{-2}$. 
The net count, {\it N.~C.}, for each source was computed as:
$$ N.~C.~(counts) = (C_{extr} - B \times \pi R_{ext}^{2}).$$
To compute the count rate, {\it C.~R.}, we also need to determine the 
enclosed fraction of the PSF, $f_{PSF}$, and the effective exposure time, 
$t_{eff}$. 
We computed $f_{PSF}$, using the CIAO tool {\it mkpsf}. 
The effective exposure time is determined from exposure maps, 
which were generated by the CIAO tool {\it mkexpmap}.
The exposure maps are images of effective area that contain information 
about instrumental artifacts that are both energy- and position-dependent.
The effective exposure time is the exposure time corrected to account 
for a 1.3\% loss due to readout, and scaled to the effective area within 
$R_{ext}$ with respect to the effective area at the optical axis.
The count rate is then computed as:
$$C.~R.~(counts/ks) = N.~C./ (f_{PSF} \times t_{eff}).$$

We carefully inspected the light curves of all the sources (see 
Sec.~\ref{variability}) and their appearance in the image of the 
Chandra field of view. 
A total of 48 (15\%) sources were rejected from the original list: 
43 sources had light curves consistent with cosmic ray afterglows,
2 sources were detected twice, since {\it wavdetect} was run separately 
in each CCD, and 3 sources had net counts less than zero. 
Of the 43 light curves having a cosmic ray shape, 40 (93\%) contain
cosmic ray afterglows flagged by the pipeline.
Our final list of 263 X-ray sources is listed in Table~\ref{tab_xray}.

\subsection{X-ray Luminosities}\label{xray_lum}

\begin{figure}
\epsscale{0.7}
\plotone{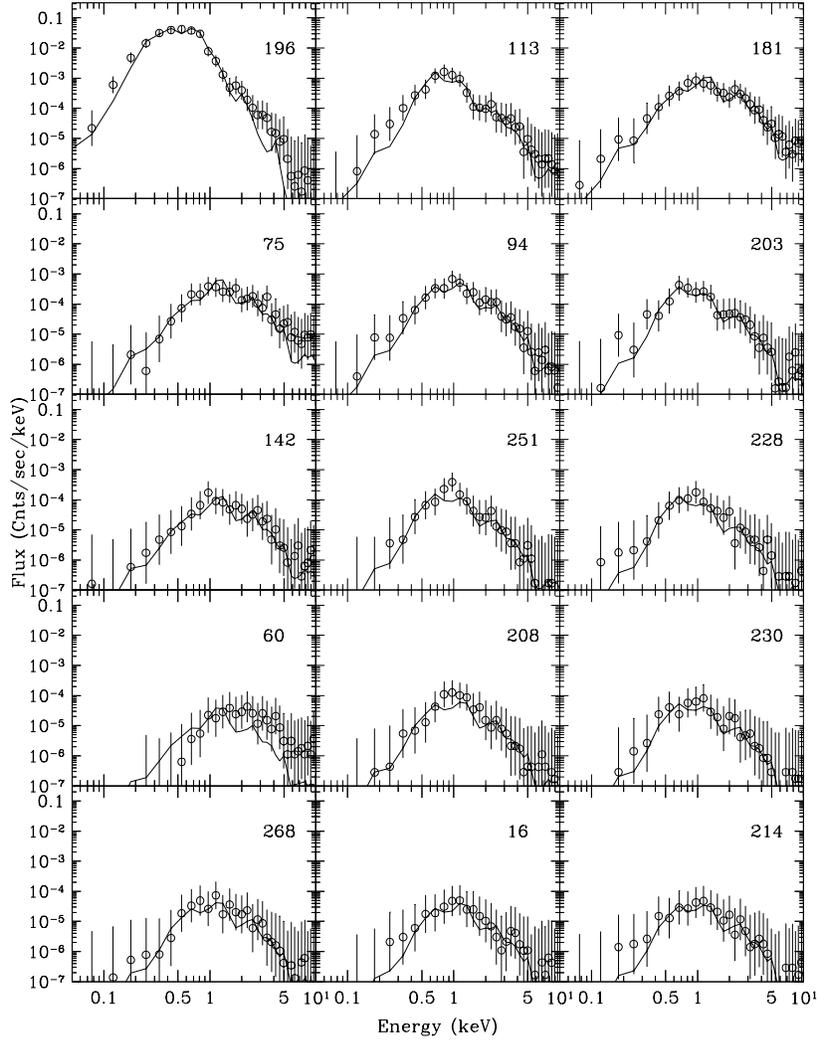}
\caption[spectra.ps]{Spectra of the 15 brightest sources in our sample.
The solid lines show the best model. 
We used a one plasma temperature model for the brightest source, 196, a
known O star, and two plasma temperature models for the rest of the
sample. Source 196 was not included in the determination of the conversion
factor that leads to \lx.
\label{spectra}}
\end{figure}

We selected all sources with more than 500 net counts, extracted
their spectra and fit them to measure their X-ray fluxes. 
A total of 15 sources meeting this criteria are listed in 
Table~\ref{tab_spectra}.
The spectra are extracted within $R_{ext}$ using the CIAO tool 
{\it dmextract}. 
The spectra of these 15 brightest X-ray sources are shown in 
Figure~\ref{spectra}.
The Redistribution Matrix Files (RMF) and Auxiliary Response Files (ARF) 
files contain instrument response information that are 
essential for the spectral fitting process.
We use the RMF files provided by the CTI corrector by \citet{tow00}. 
ARF files were created for each source, using the CIAO tool {\it mkarf}. 
The computed ARFs were later corrected to account for the
ACIS low energy quantum efficiency degradation, using
the ACISABS absorption 
model.\footnote{http://cxc.harvard.edu/cal/Acis/Cal\_prods/qeDeg/\#Obtaining the corrarf and ACISABS tools}
We used the CIAO tool {\it sherpa} for the spectral fitting.
We adopted a photoelectric absorption model 
($xswabs$), which uses Wisconsin cross sections from \citet{mor83}.
This model has one parameter that is the equivalent hydrogen column 
density ($nH$). 
The hydrogen column density was fixed to a value of 0.08$\times 10^{22}
{\rm cm^{-2}}$ to match an extinction value of $A_V$ = 0.41, 
which is the most likely value of the observed extinction towards
NGC 2264, although the $A_V$ can be as high as 3.5 mag 
\citep[see ][for more discussion]{reb02}.
The value of the column density was derived using the relationship of
$nH = 2 \times 10^{21} A_V$.
If we consider a $nH=0.65\times 10^{22}{\rm cm^{-2}}$, which corresponds 
to three magnitudes more than the assumed $A_V$, the derived \lx~ 
would be only 0.1 dex less.
Thus, the error in \lx~ that comes from fixing $nH$ is negligible. 
We also adopted a thermal emission model ($xsmekal$), which
is based on model calculations of \citet{mew85,mew86}, and \citet{kaa92} 
with Fe L calculations by \citet{lie95}.
This model includes line emissions from several elements.
The remaining two parameters in the model - the plasma temperature 
(kT) and a normalization factor - were varied in order to fit the spectra.

We initially tried one temperature models, but they failed to fit the 
low energy part of the spectrum. 
A two temperature model significantly improved the spectral fitting 
for all the sources.
The brightest X-ray source in the sample 196 (S Mon) has a 
distinctive soft X-ray spectrum, characteristic of its early spectral
type (see Sec.~\ref{ostar}). 
It also has a high count rate of 155 counts ks$^{-1}$,
and may be affected by pileup (at high count rate the observed
flux is no longer a linear function of the count rate).
Given the spectral characteristics of S Mon and its possible pileup
effects, we did not include it in the determination of the conversion
factor that leads to the determination of X-ray luminosity for
our sample of X-ray sources. 
The spectral parameters obtained from the two plasma temperature models
are listed in Table ~\ref{tab_spectra}.

The X-ray flux of S Mon listed in Table ~\ref{tab_spectra} is the 
integration of the two temperature model between 0.3 and 8.0 keV.
The X-ray flux for the rest of the bright sources is determined from 
the best spectral model derived from the mean model parameters. 
We computed mean plasma temperatures of (0.51$\pm$0.06) keV and
(2.5$\pm$0.3) keV. 
The mean plasma temperatures are held constant and the integration of
the best fit between 0.3 and 8.0 keV provides the X-ray flux.
The resulting X-ray fluxes are listed in Table ~\ref{tab_spectra}.

\begin{figure}
\epsscale{0.7}
\plotone{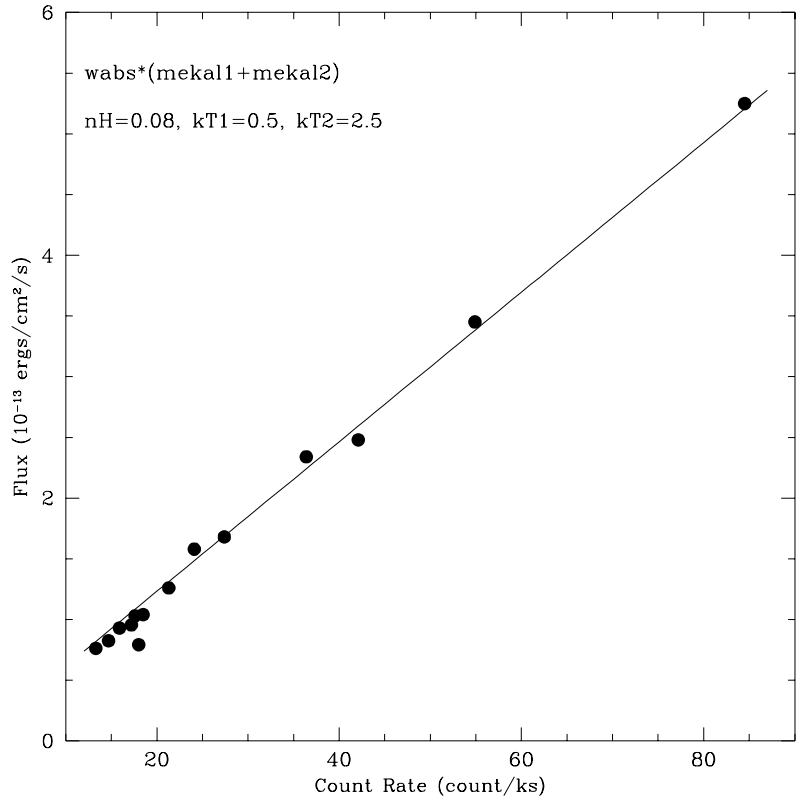}
\caption[cfactor.ps]{Count Rate versus Flux for the 14 bright X-ray sources
with two plasma temperature models
used to determine a conversion factor.
The solid line shows the determined conversion factor of
(6.16$\pm$0.13)$\times10^{-15}$ (erg/cm$^{2}$/s)/(counts/ks).
\label{cfactor}}
\end{figure}

We use the X-ray fluxes of the 14 bright sources to compute a X-ray 
flux weighted conversion factor between count rate and X-ray flux. 
The obtained conversion factor is (6.16$\pm$0.13)$\times10^{-15}$
(erg/cm$^{2}$/s)/(counts/ks). 
Figure ~\ref{cfactor} shows the relationship between $C. R.$
and X-ray flux for the 14 X-ray sources used to determine the conversion 
factor. The solid line corresponds to the derived conversion factor.
We compiled a list of conversion factors obtained by published 
studies using ACIS-I data in young stars; see Table~\ref{tab_compare}.
The values we list for \citet{fei02} and \citet{get02} were computed
using the count rate and X-ray luminosity they provide in their tables 
3 and 1 respectively.
Our conversion factor is in good agreement with the ones
derived by \citet{kri01} and \citet{get02}, and it differs 
from the other values by a factor of less than 2. 
The X-ray flux for our catalog of X-ray sources in NGC 2264,
listed in column 10 of Table~\ref{tab_xray}, was 
computed using the derived conversion factor.
The X-ray luminosity,~\lx, listed in column 11 of Table~\ref{tab_xray}, 
is computed assuming a distance to NGC 2264 of 760 pc \citep{sun97}.

\begin{figure}
\epsscale{0.7}
\plotone{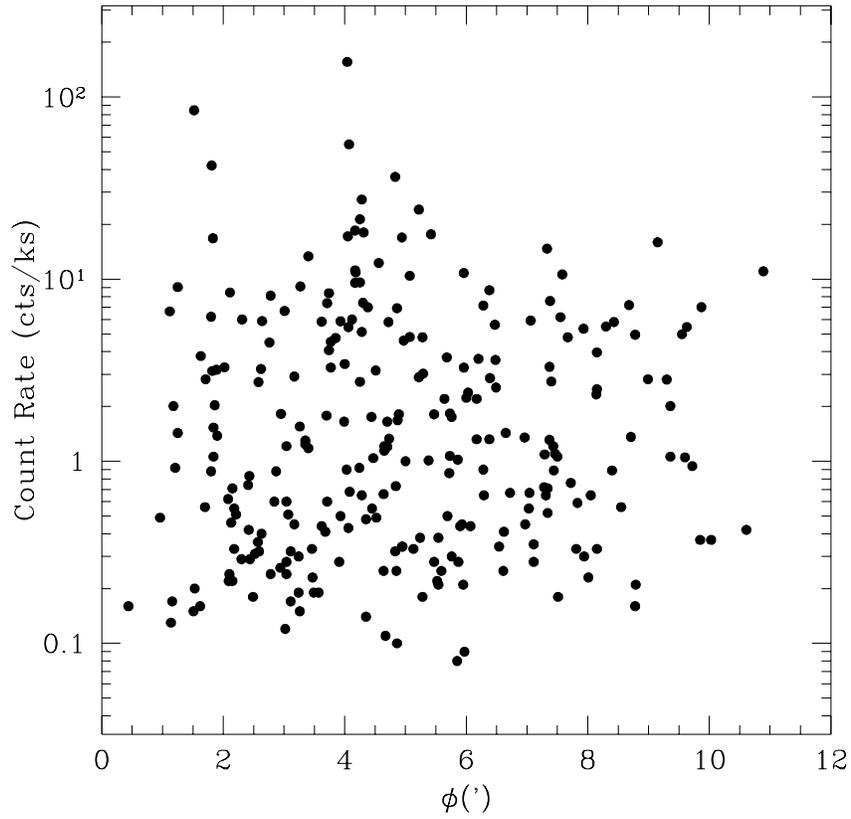}
\caption[limit.ps]{Count Rate of detected X-ray sources as a function
of the off-axis angle ($\phi$). The limiting count rate of our observations
varies with $\phi$, since the PSF of the Chandra observations varies
across the field of view.
\label{limit}}
\end{figure}

The limiting luminosity in our X-ray observations varies within the 
field of view, because of the variation of the PSF across the field. 
In Figure~\ref{limit}, we have plotted the Net Rate of the detected
X-ray sources as a function of the off-axis angle, $\phi$. 
The faintest source is located at $\phi \sim 6 \arcmin$ and
it has a count rate of 0.08 counts/ks, corresponding to a X-ray 
luminosity of 28.5 at the distance of NGC 2264. 
About 80 \% of our sources are located within $\phi = 7\arcmin$.
At that off-axis angle the limiting count rate has increased to
0.12 counts/ks, corresponding to a X-ray luminosity of log(\lx)=28.7 
at the distance of NGC 2264. 
At $\phi = 10\arcmin$, we cannot detected X-ray sources fainter than 
0.35 counts/ks (log(\lx) = 29.2).
Therefore, we adopt a value of log(\lx)=28.5 dex as the limiting 
X-ray luminosity for our observations, keeping in mind that this
value holds for sources located within $\phi \sim 6 \arcmin$.

\begin{figure}
\epsscale{0.7}
\plotone{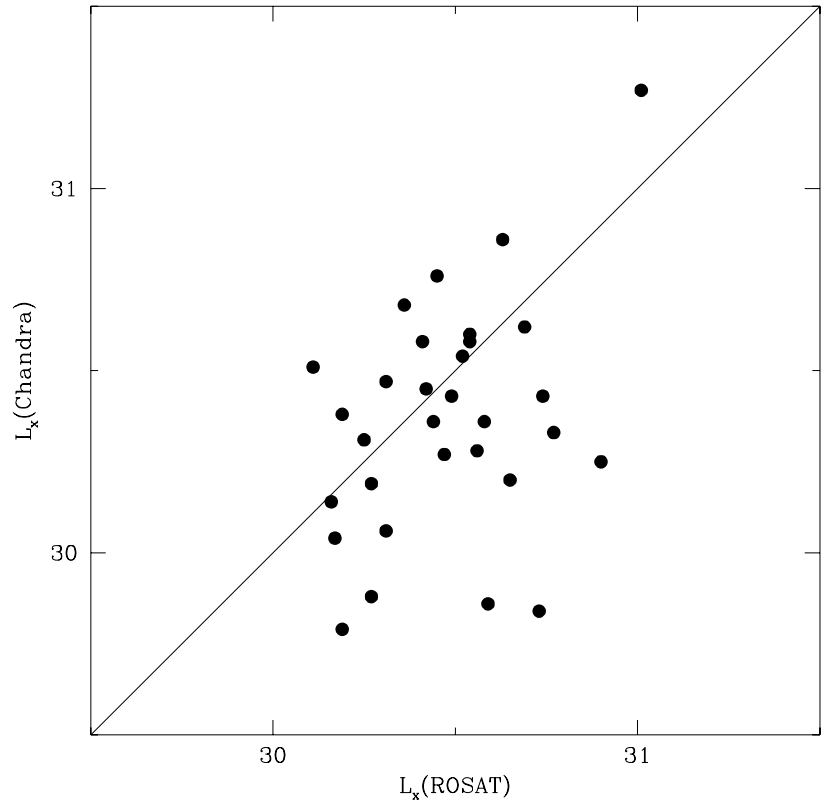}
\caption[rosat.ps]{Our X-ray luminosities (from Chandra) compared to
the ones from \citet{fla00} (from ROSAT). 
The solid line represent the equality of both measurements. 
Our observations are well-matched to those from \citet{fla00}.
\label{rosat}}
\end{figure}

 
\citet{fla00} observed a 40'$\times$100' field towards NGC2 2264
using ROSAT, detecting 169 sources. Our Chandra field (17'$\times$17')
is covered within the spatial extent of the ROSAT
observations but goes 1.5 dex deeper in log(\lx) 
(a factor of $\sim$ 30 in \lx).
There are 34 sources in common between our catalog of X-ray
sources in NGC 2264 and the ROSAT sample of \citet{fla00}. 
We compared our X-ray luminosities with the ones given by
\citet{fla00} (see  Figure~\ref{rosat}). 
We found a mean difference of --0.1 dex with a standard deviation
of 0.3 dex.
This difference does not include an offset of +0.3 dex in the
ROSAT \lx~, as described in \citet{fla03}.
Given that many of these PMS stars are highly variable (see 
Sec.~\ref{variability}), we conclude that our luminosities are 
consistent with respect to those of \citet{fla00}.

\subsection{Variability}\label{variability}
 
\begin{figure}
\epsscale{0.7}
\plotone{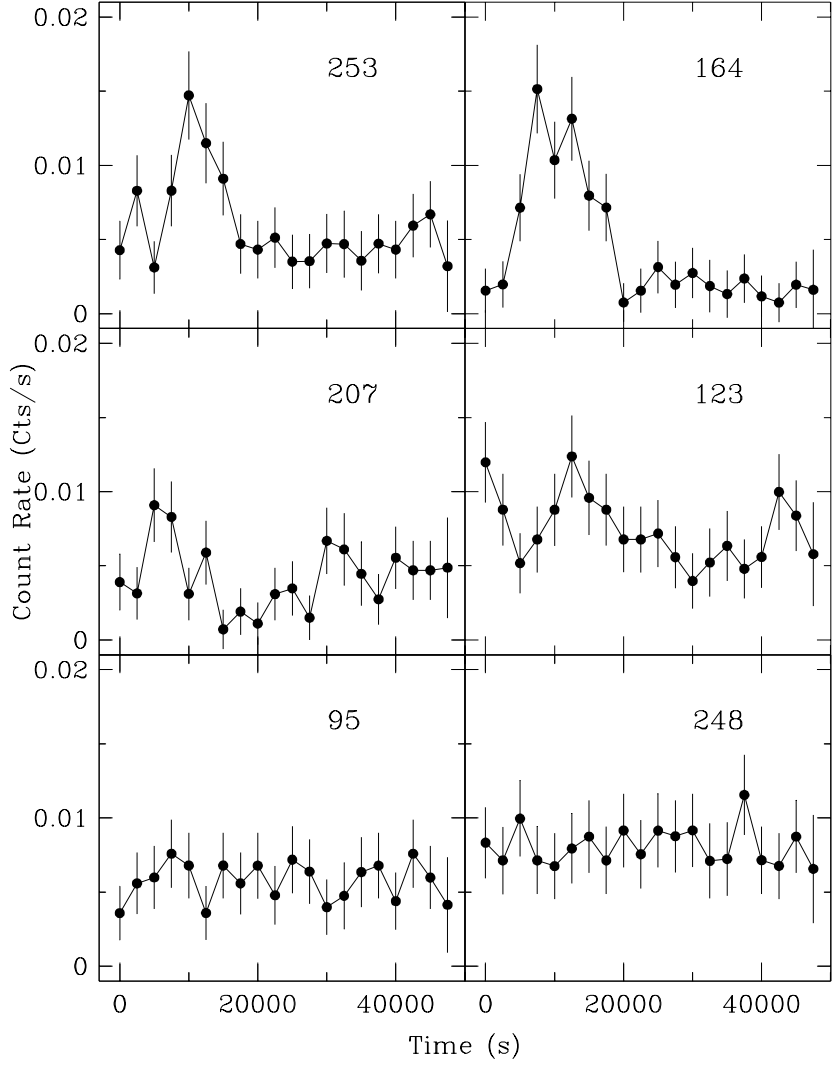}
\caption[light_curves.ps]{Examples of light curves for six X-ray sources
of comparable luminosity from our sample. 
The two sources in the top panels show flare-like light curves. 
The two sources on the middle panels are variable, and 
the two sources in the bottom panels are constant.
There are 41 variable sources in our sample of 263 X-ray objects; 14 of
them show a flare-like light curve. 
\label{light_curves}}
\end{figure}

Light curves were determined for all 313 sources detected
by {\it wavdetect} using the CIAO tool {\it lightcurves}
with a bin time of 2500 s.
The statistics of the light curves of the sources of our 
X-ray catalog were obtained using the Xronos
\footnote{http://heasarc.gsfc.nasa.gov/docs/xanadu/xronos/xronos.html} 
tool {\it lcstats}. 
This provides, among other values, the constant source probability 
as derived from the Chi-square value comparing the data with a 
constant light curve, $P_{c}(\chi^{2})$.
The $P_{c}(\chi^{2})$ values are listed in column 12 of Table~\ref{tab_xray}.
The light curves of the sources with $P_{c}(\chi^{2}) < $ 90\% were 
analyzed further, since they are the most likely to be variable.
We obtained the reduced $\chi^{2}$ of those sources, using two additional 
light curves with bin times of 5000 s and 7500 s.
If the reduced $\chi^{2}$ obtained for both bin times was
less than 2.5, then we consider those sources to have constant
light curves.
Then, we defined a variable source as those having 2500 s bin light curves 
with $P_{c}(\chi^{2}) < $ 90\%, and 5000 s and 7500 s bin light curves
with reduced $\chi^{2} > $ 2.5.
There are 41 variable sources that meet this criteria.
Variable sources are marked with a 'v' in column 13 of 
Table~\ref{tab_xray}.
Among the variable sources, ten have known periods from \citet{mak04},
but we find no obvious dependence of X-ray flux with rotational phase for
9 of these stars (see bellow the discussion about source 181).
There are 14 variable sources which show a flare shape, defined as a 
rapid increase and a slow decrease in the X-ray flux.  
Variable sources with a flare-like light curve are
marked with an additional 'f' in column 13 of Table ~\ref{tab_xray}. 
In Figure ~\ref{light_curves}, we show a selection of light curves
of sources of comparable luminosity.
In the top panels, we show two light curves with a flare shape;
in the middle panels, there are two variable light curves.
Finally in the bottom panels, two constant light curves are plotted.
There are six sources that show a possible flare pattern, described 
as an increase in X-ray flux happening at the end of our observations
or a decrease in X-ray flux occurring at the beginning of our observations.
Variable sources with a possible flare pattern are marked
with an additional 'p' in column 13 of Table~\ref{tab_xray}. 
\citet{gag95} detected 10 flaring objects in Orion using ROSAT observations, 
over a sample of 389 X-ray sources ($\sim$ 3\% of flaring objects);
\citet{pre02} detected 14 flaring objects in IC 348 using Chandra 
observations 
over a sample of 215 X-ray sources ($\sim$ 7 \% of flaring objects);
we detected a comparable fraction , about 8\% flaring sources (5\% excluding
possible flares).

\begin{figure}
\epsscale{0.7}
\plotone{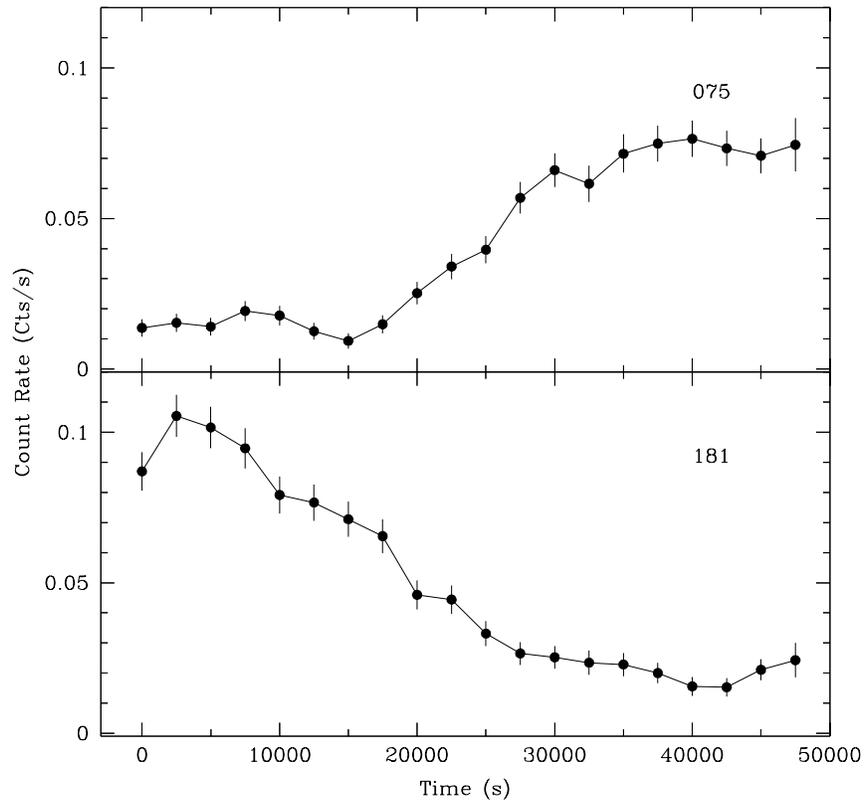}
\caption[light_curves_2.ps]{Light curves for three X-ray sources
that show a steady increase or decrease of flux in their light curves.
The amount of variation is about a factor of 10 over about 10 hours.
\label{light_curves_2}}
\end{figure}

There are two additional variable sources which show a distinctive 
pattern of variation,  either a steady increase or a steady decrease 
in X-ray flux. 
The sources that show this pattern are 75 (R 2752) and 181 
(R 3245), they are marked with an additional 's' in column 13 of 
Table~\ref{tab_xray}, and their light curves are shown in 
Figure~\ref{light_curves_2}.
The amount of variation in these two cases is about a factor of
10 in about 10 hours. 
A similar pattern is observed in the light curves of three X-ray
IC 348 sources determined by \citet{pre02} using Chandra observations.
They mention that those light curves may be explained by rotational
variability, as described by \citet{ste99}, who present a model of
rotational modulation of X-ray flares in three T Tauri stars and
Algol.
The amplitude of the variations modeled by \citet{ste99} vary from 
a factor of 2.5 to a factor of 20, depending on the strength of
the X-ray flare.
Although both X-ray sources 181 and 75 have optical counterparts, 
only 181 (R 3245) has an observed period of 1.21 days (29 hrs) 
from \citet{mak04}.
If the lightcurve of 181 can be understood by rotational modulation 
of X-ray flares, then the observed decay time ($\sim$ 11 hrs) should 
be less than half of the stellar period. 
Therefore, the period implied by the X-ray light curve should be greater 
than 22 hrs, which is consistent with the optical observations.
 
\section{DISCUSSION}

\subsection{Description of the optical/infrared catalog}\label{catalog}
 
We correlated the X-ray sources with a catalog of stars in NGC2264;
see Table~\ref{tab_optical}.
We constructed the catalog by combining our original $UBVRI$ survey from
\citet{reb02} (``R'' names) with other published catalogs.  
If optical photometry and/or spectral types were not available in 
\citet{reb02}, we took photometry from \citet{par00}, \citet{fla99}, 
or \citet{sun97}, in that order, augmented by 15 spectral types from the 
SIMBAD database.  
All of the $JHK$ photometry comes from 2MASS, with the exception of a
handful of stars slightly fainter than the 2MASS limit; those $JHK$
values come from \citet{reb02}.

We included several other catalogs primarily to keep track of
commonly-used nomenclature; see Table~\ref{tab_xids}.  
\citet{her54}, updated by \citet{mar80}, lists spectral types and 
is the origin of the ``LkHa'' nomenclature; 
\citet{wal56} provides spectral types, membership probabilities, and the
commonly-used Walker (or ``W'') names; \citet{ogu84} conducted an H$\alpha$
survey to identify young stars and provides some spectral types; and 
\citet{vas65} and \citet{her88} are the origin of the ``VSB'' and 
``HBC'' names, respectively.
\citet{fla00} conducted a ROSAT survey of this cluster, and we
included these data in our catalog as well (see Sec.~\ref{xray_lum} above).  
\citet{fla00} ROSAT sources are named as ``FX''.
Because we are also interested in rotational information (for our future
work), we included in 
our catalog periods from \citet{mak04} and \citet{kea98}, as well as
projected rotational velocities ($v \sin i$) from 
\citet{mcn90}, \citet{sod99} and \citet{ham03}.

We found that 213 (81\%) of our 263 X-ray sources have optical and/or
infrared counterparts (108 (41\%) have spectral types and 44 (17\%)
have known periods).  
Those 213 sources and their corresponding optical and/or infrared 
photometry are listed in Table~\ref{tab_optical}. 
Most, but probably not all, of these 213 stars are likely to be
members of NGC2264.
Other names of the sources given by the different catalogs used in our
compilation are listed in Table~\ref{tab_xids}. 

\begin{figure}
\epsscale{0.7}
\plotone{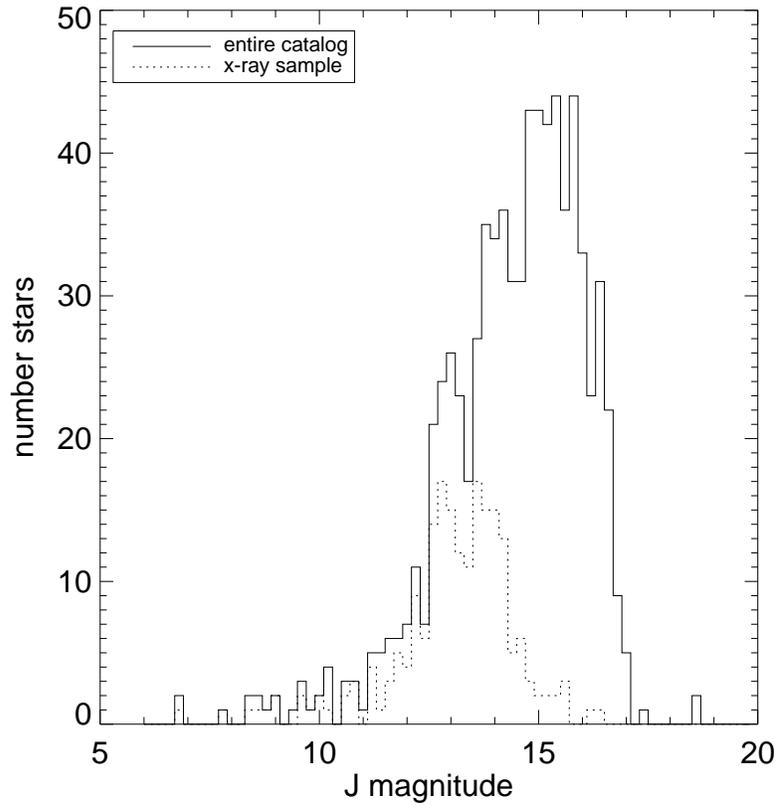}
\caption[histogram.ps]{$J$ magnitude histogram of all the sources
with $J$, $H$, and $K$ magnitudes in our Chandra field.
The solid line corresponds to all the sources, and the dotted line
corresponds to all the sources with X-ray counterparts.
All the X-ray sources should have been matched to sources in our catalog
if they are associated with stars earlier than M0 at the distance of NGC
2264.
\label{histogram}}
\end{figure}

There are 747 stars in our optical/infrared catalog with 
$J$, $H$, and $K$ photometry and with positions inside the field 
of view of the Chandra observations. 
Among those 747 stars, 199 (27\%) have X-ray Chandra counterparts.
Figure~\ref{histogram} shows a $J$ magnitude histogram of all 
the 747 stars with $J$, $H$, and $K$ photometry and positions 
inside the Chandra field (solid line) and the histogram 
of the 199 X-ray Chandra counterparts (dotted line).
The completeness limit of the infrared sample is essentially that of 2MASS.
We can see in Figure~\ref{histogram}
that our infrared sample goes deeper than the sample of stars
with X-ray counterparts. 
A similar behavior is seen in the optical sample histogram. 
This means that all the X-ray sources should have been matched to sources 
in our optical/infrared catalog if they are associated with stars
earlier than M0 at the distance of NGC 2264.
There are 51 X-ray sources that do not have optical nor infrared 
counterparts. 
Based on the limiting magnitude of our optical catalog
($V\sim$ 19 mag), the X-ray to optical flux ratio ,$f_X/f_V$,
of the sources with only X-ray detection
is greater than about $10^{-2.5}$. 
This means that these sources could be either M stars or 
extragalactic objects \citep{sto91}.
NGC 2264 is located 2 degrees above the galactic plane, roughly
towards near the anticenter. Therefore, most of the M stars 
should be foreground, and hence detected by 2MASS.
Considering the presence of a dark cloud behind NGC 2264, 
the detection of background M stars in X-rays is unlikely.
Source counts in the Chandra South and North Deep Field
\citep{ros02,bra01} predict the presence of about 100 AGN in the
ACIS field of view at the flux limit of our observations. 
Thus, we believe that the most likely explanation is that the 51 
objects detected only in X-rays are active galaxies.

\subsection{Color-Color diagram.}

\begin{figure}
\epsscale{0.7}
\plotone{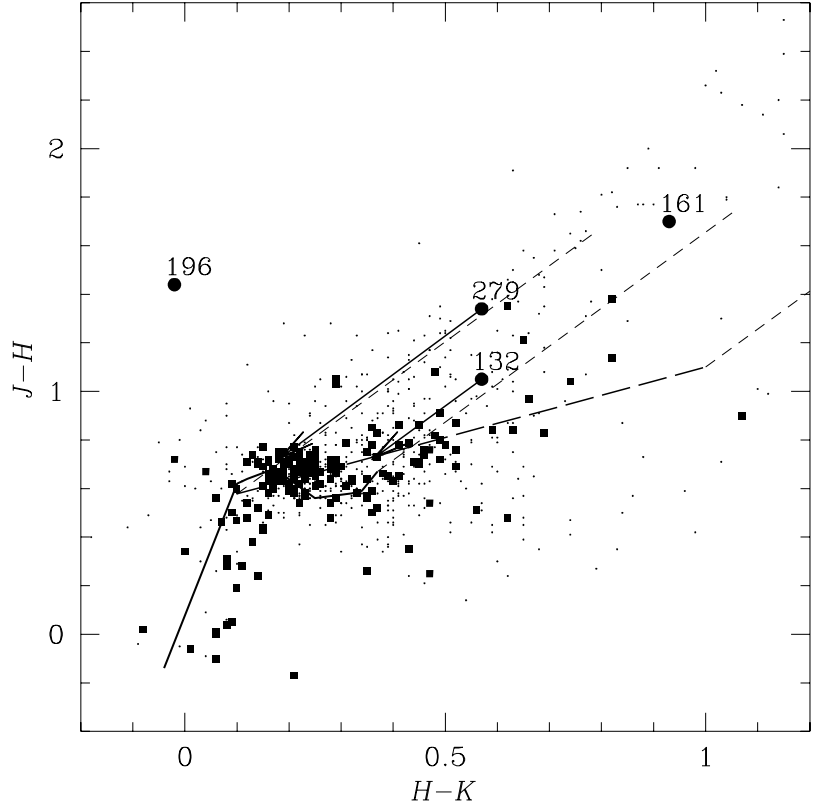}
\caption[color_color.ps]{Observed infrared color-color diagram of sources 
located in the field of the ACIS camera. The $filled~squares$ denote stars 
with X-ray counterparts, and the $dots$ mark the position of stars 
without X-ray counterparts.
The $thick~line$ marks the location of main sequence colors, 
the $dotted~line$ the locus of CTTs from \citet{mey97},
and the $dashed~lines$ are reddening vectors; the length corresponds
to $A_V = 10$ magnitudes.
Source 196 is an O star with a suspect J magnitude and source 161 is a
galaxy; see text.
Sources 279 and 132 are also discussed in Figure~\ref{cmd}. 
\label{color_color}}
\end{figure}

In Figure~\ref{color_color}, we have plotted the $J-H$, $H-K$
color-color diagram of all the infrared sources in the field
of view of our Chandra observation. 
Most of the X-ray sources with infrared colors are located near
the locus of the classical T-Tauri (CTT) stars \citep{mey97}. 
There are two X-ray sources that do not follow the position of the 
rest of the sources. They are sources 161 and S Mon (196) and they
are marked in the color-color diagram with their X-ray identification 
numbers. 
S Mon (see Sec.~\ref{ostar}, is a O star;
its position in Figure~\ref{color_color} is correct in
$H-K$, but its 2MASS $J$ magnitude is suspect and can be explained
by contaminated from a nearby diffraction spike.
The second source, 161, is a field galaxy (see Sec.~\ref{galaxy}).

\subsection{Color-Magnitude Diagrams.}

\begin{figure}
\epsscale{0.7}
\plotone{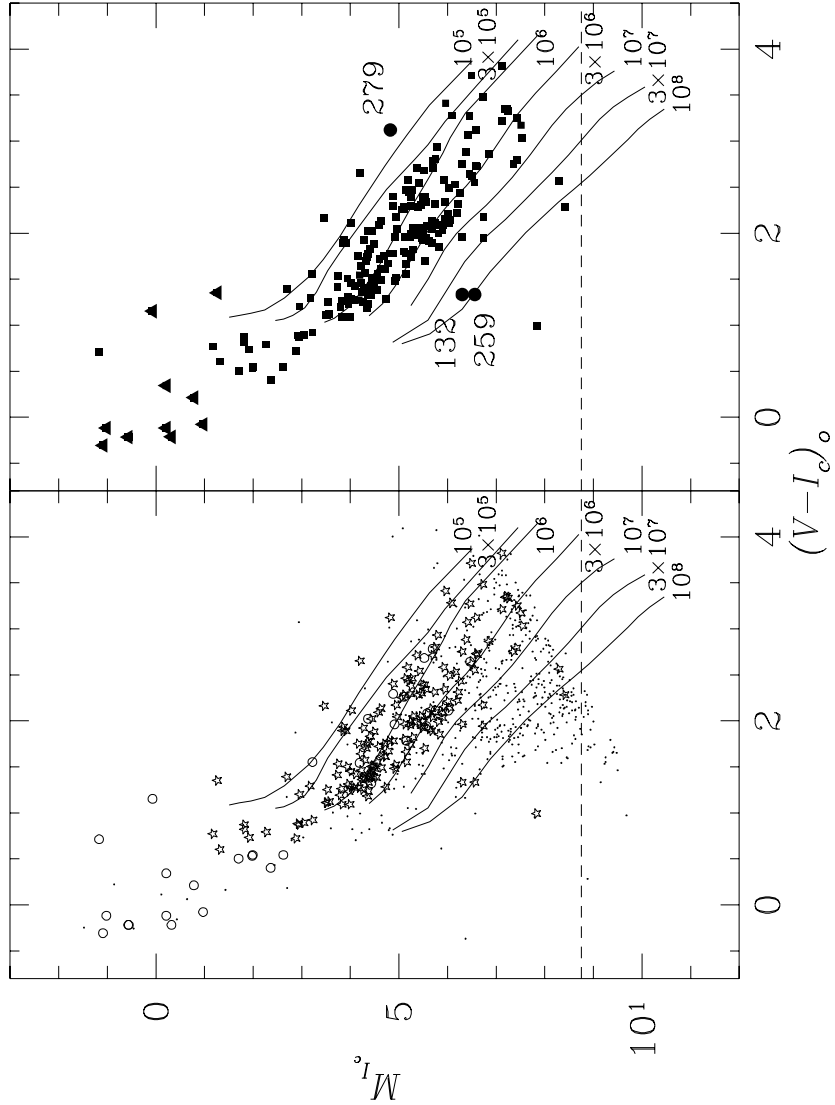}
\caption[cmd.ps]{Dereddened optical color magnitude diagrams of sources located 
in the field of the ACIS camera. 
In the left panel, $open~circles$ denote stars with X-ray counterparts 
and ${\rm L_x/L_{bol} < 10^{-4}}$, $stars$ denote sources with X-ray 
counterparts and ${\rm L_x/L_{bol} \geq 10^{-4}}$ , and the $dots$ mark 
the position of stars without X-ray counterparts.
In the right panel, only optical sources with X-ray counterparts are plotted.
Early type stars are marked with $filled-triangles$ and sources discussed in 
the text are marked with $filled~circles$ and their respective X-ray numbers.
The dashed line corresponds to $M_{I_{c}}$ = 8.5 mag., which is the lower 
limit for a low mass star rotating at the saturation level.
The solid lines denote isochrones from \citet{dan98}.
Most of our X-ray sources are $<$ 3 Myr old and they are clearly younger
than the general population.
\label{cmd}}
\end{figure}

There are 686 stars in our optical/infrared catalog with
$I_c$ and $V$ photometry and with positions inside the field
of view of the Chandra observations.
Among those 686 stars, 201 have X-ray Chandra counterparts.
In Figure~\ref{cmd}, we have plotted ($V-I_c$)--$M_{I_{c}}$
color magnitude diagrams of the optical sources in the field 
of view of our Chandra observations. 
The dereddened $V-I_c$ color and the absolute magnitude $M_{I_{c}}$ were 
obtained assuming an average extinction of $A_V$ = 0.41, and
dereddening relationships from \citet{fit99} and \citet{mat90}.
In both panels, we have also plotted isochrones of from \citet{dan98},
as a reference.
The dashed line corresponds to $M_{I_{c}}$ = 8.75 mag., which
is the lower limit for a low mass star rotating at the saturation 
level (\lx/\lbol=--3) with \lx=28.5 (limit luminosity of our X-ray sample). 
This means that the most slowly rotating stars with optical counterparts
should be detected by our X-ray sample.
For comparison, the corresponding line from the ROSAT \citet{fla00} study
would be at $M_{I_{c}}$ = 5 mag.

\begin{figure}
\epsscale{0.7}
\plotone{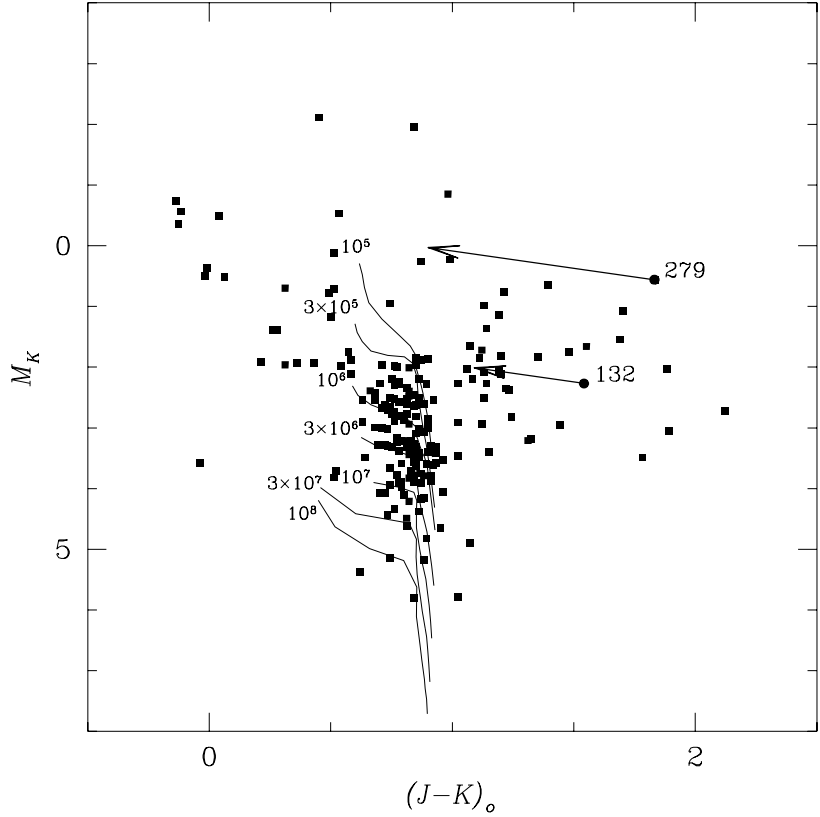}
\caption[cmd3.ps]{Dereddened infrared color magnitude diagram of sources located
in the field of the ACIS camera. Only infrared sources with X-ray counterparts 
are plotted. Sources 132 and 279 are marked with their X-ray number, and
their dereddened positions are marked with arrows.
The solid lines denote isochrones from \citet{dan98}.
Sources 279 and 132 are likely to be the youngest and most embedded of our
X-ray sources; the unusual optical colors of source 132 could be explained
by scattered light from an edge-on disk.
\label{cmd3}}
\end{figure}

Most of our X-ray sources are younger than 3$\times 10^{6}$ years
at the distance of NGC 2264, and furthermore the X-ray sources are
heavily concentrated towards the youngest isochrones with respect to
the general population. 
All the sources with $M_{I_{c}}$ brighter than 0.1 magnitudes have
\lx/\lbol~ $< 10^{-4}$. Most of those sources have spectral
types earlier than A0 (see right panel of Figure~\ref{cmd}). 
There are 5 X-ray sources that appear to be much older than $10^{7}$ years, of
which one of them (12) is not a member of NGC 2264, two
of them (146 \& 273) lack proper motion membership information, and 
two of them (132 \& 259) are probably members
\citep[using preliminary membership information as discussed in ][]{reb02}.
Unfortunately, the optical and infrared photometry of source 259 is 
unreliable, because it is contaminated by a saturated column 
from a nearby bright star.
Source 132 is classified as a M0 star.
There are four sources that are located above the $10^{5}$ year old isochrone,
of which two of them (75 \& 136) are not members of NGC 2264, one of them
(60) lacks proper motion membership information, but it is an A0 star, and one
of them (279) is  probably a member 
\citep[using preliminary membership information as discussed in ][]{reb02}.
Source 279 is classified as a K5 star.
We have dereddened sources 279 and 132, using the intrinsic $J-H$, $H-K$ and 
$R-I_c$ 
colors of stars of their respective spectral types. 
For source 279, we derive a $A_V$ value of 5.7 magnitudes and for 
source 132, $A_V$ is about 3.0 magnitudes. 
The locations of sources 279 and 132 are indicated along with their
dereddened positions in Figure~\ref{color_color}.
The dereddened colors of source 279 are consistent with colors of a K star, 
similar to its spectral type, and the dereddened colors of source 132 fall 
in the locus of CTT stars.
In Figure~\ref{cmd3}, we have plotted the $(J-K)-M_K$ color magnitude
diagram of all the infrared sources with X-ray counterparts, with 
isochrones from \citet{dan98}. 
The location of sources 279 and 132 is indicated
along with their dereddened position.
It is possible that these two objects are the youngest and most embedded of
our X-ray sources. The unusual optical colors of source 132 could be explained
by scattered light from an edge on disk.

\subsection{Upper Limits.}

There are 16 stars for which rotation and spectral type 
information exist in our optical/infrared catalog, and
for which no X-ray counterpart was found.
In order to allow us to use these stars in the next paper, 
we determined upper limits 
for their X-ray luminosity in the following way. 
We extracted the counts with a $R_{extr}$ equal to the 
95\% encircled energy radius. We computed the background
counts the same way as our detected sources.
We used the Bayesian statistics method from \citet{kra91}
to determine the net count upper limit to a confidence
level of 95\%. Then, we computed $f_{PSF}$ and $t_{eff}$ in
the same manner as our detected sources, to determine 
the upper limit for the count rate. The upper limit for
the X-ray flux was determined using our derived conversion factor,
and the upper limit for the X-ray luminosity was computed
assuming a distance to NGC 2264 of 760 pc \citep{sun97}.
The upper limit results are listed in Table~\ref{tab_upp_lim}.

\subsection{Other Interesting Objects.}

\subsubsection{S Mon}\label{ostar}

X-ray source 196 is associated with S Mon, a O7 V star.
The X-ray spectrum of S Mon is well fitted by two plasma temperatures
of 0.18 keV and 0.57 keV, but nearly 80\% of the total X-ray flux comes
from the 0.18 keV component. We derive a log(\lx/\lbol)=--6.78, which 
is in good agreement with the canonical value of log(\lx/\lbol) $\sim$--7
observed in O stars \citep{chl89,cas94}.
\citet{ber96} provide a catalog of OB stars observed in the ROSAT All-Sky 
Survey, which includes S Mon. They derive (from hardness ratios) a 
X-ray plasma temperature of 0.23 keV and log(\lx/\lbol)=--6.63, 
which are in good agreement with the values we derive considering the 
differences in the applied techniques (spectral fitting vs. hardness ratios).
The X-ray emission mechanism for early type stars is thought to
be very different from the magnetic reconnection flares 
that generate the X-rays in low mass young stars. 
Models predict that the X-ray emission of hot stars comes from shocks 
formed within a radiatively driven wind 
\citep{luc80,luc82}, producing X-rays of about 0.5 keV.
The shocks may be formed by line-forced instabilities \citep{owo88,fel97}.
Evidence coming from grating X-ray spectroscopy suggests that magnetically
confined hot plasma near the surface of the star may be important
in some early type stars that show an unusually hard X-ray spectrum 
(plasma temperature higher than 2 keV)
and narrow X-ray emission lines \citep[e.g. ][]{bab97,coh97,coh03}.
The lack of a hard component in the X-ray spectrum of S Mon may suggest
that its X-ray emission is likely produced by shocks within the radiatively driven stellar wind.

\subsubsection{Walker 90}

Walker 90 is an early-type emission-line star that lies below the
zero-age main sequence in the $V$, $B-V$ color magnitude diagram 
\citep{str71}, suggesting very non selective extinction. 
\citet{sit84} confirmed the existence of anomalous circumstellar
dust extinction, which they explained by a graphite-silicate
mixture with larger grains than those present in the
diffuse interstellar medium.
Walker 90 is in our Chandra field, but no X-ray source was
detected at its position. 
An upper limit for its X-ray luminosity is provided in 
Table~\ref{tab_upp_lim}.

\subsubsection{A Relatively Bright Active Galaxy towards Our 
Chandra Field}\label{galaxy}

Our X-ray source 161 is coincident in position and 
has similar morphology to
the extended infrared source 2MASX J06405286+0948570.
The $J$, $H$, and $K$ magnitudes from the 2MASS extended
source catalog are listed in Table~\ref{tab_optical}.
This source is also a known radio source, with a 
flux of 6.45 mJy/beam at 3.6 cm \citep{rod94}
and a flux of 61.9 mJy at 20 cm \citep{con98}.
Given its radio and X-ray emission, we conclude that
source 161 is a radio-quiet AGN \citep{elv94},
seen though the molecular cloud.

\section{CONCLUSIONS}

We present a catalog of NGC 2264 X-ray sources.
The observations were taken with the ACIS-I on board the
Chandra X-ray Observatory.
The catalog, consisting of 263 sources, includes X-ray luminosity, 
optical and infrared photometry
and X-ray variability information. We found 41 variable sources,  
14 of which have a flare like light curve, and 2 of which
have a pattern of a steady increase or decrease. 
From the optical and infrared counterparts of the X-ray
sources, we have learned that most of the X-ray sources have 
colors consistent with CTTs that are younger than 
3$\times 10^{6}$ years. 

This catalog of X-ray sources will
be used to study the relationship between rotational properties
and X-ray characteristics of NGC 2264 stars in paper II
\citep{reb03}.
We plan to discuss correlations of \lx/\lbol~ with
rotation rate (period and $v$sin$i$), disk indicators ($I_c-K$,
$H-K$, $U-V$, and $H\alpha$), and mass accretion rate as derived
from $U-V$ excess.  We will also compare the \lx/\lbol~ values
found here with those from other young clusters.

\acknowledgements
SVR gratefully thanks Jeonghee Rho, Kenji Hamaguchi, Ettore Flaccomio, 
Peter Freeman \& Scott Wolk for the useful correspondence about
X-ray and Chandra data processing and analysis.
SVR also thanks August Muench, Paul Eskridge, Richard Pogge,
Phil Appleton, Mark Lacy, and Dario Fadda for interesting discussions.
We thank the anonymous referee for her/his careful review of the
manuscript.
Financial support for this work was provided by NASA grant GO2-3011X.
This research has made extensive use of NASA's Astrophysics Data System
Abstract Service, the SIMBAD database, operated at CDS, Strasbourg, 
France, and the NASA/IPAC Infrared Science Archive, which is operated 
by the Jet Propulsion Laboratory, California Institute of Technology, 
under contract with the National Aeronautics and Space Administration.
The research described in this paper was partially carried out at
the Jet Propulsion Laboratory, California Institute of Technology,
under a contract with the National Aeronautics and Space Administration.

\clearpage


\clearpage

\clearpage

\clearpage

\clearpage

\clearpage

\clearpage

\clearpage

\clearpage

\clearpage

\clearpage

\clearpage

\clearpage

%
%
\clearpage
\begin{deluxetable}{rcllcccccccrc}
\tabletypesize{\scriptsize}
\rotate
\tablenum{1}
\tablewidth{0pt}
\tablecaption{The Catalog of X-ray Sources.\tablenotemark{a} \label{tab_xray}}
\tablehead{
\colhead{X-ID} & 
\colhead{Name} &
\colhead{RA} &
\colhead{DEC} & 
\colhead{$\phi$} &
\colhead{$R_{ext}$} &
\colhead{$f_{PSF}$} &
\colhead{$t_{eff}$} &
\colhead{$C.~R.$} &
\colhead{Flux} &
\colhead{log(\lx)} &
\colhead{$P_{c}(\chi^{2})$} &
\colhead{Comments\tablenotemark{b}} \\ 
\colhead{} &
\colhead{} &
\colhead{(2000)} &
\colhead{(2000)} &
\colhead{(')} &
\colhead{(")} &
\colhead{} &
\colhead{(ks)} &
\colhead{(count/ks)} &
\colhead{(erg/cm$^{2}$/s)} &
\colhead{log(erg/s)} &
\colhead{\%} &
\colhead{} \\
\colhead{(1)} &
\colhead{(2)} &
\colhead{(3)} &
\colhead{(4)} &
\colhead{(5)} &
\colhead{(6)} &
\colhead{(7)} &
\colhead{(8)} &
\colhead{(9)} &
\colhead{(10)} &
\colhead{(11)} &
\colhead{(12)} &
\colhead{(13)}
}
\startdata 
  1 & CXORRS J064009.6+095338 & 6 40  9.69 & 9 53 38.355 &  9.85 & 14.07 &  0.99 & 40.5 &   0.37 &  0.228E-14 & 29.20 & 100 &     \\
  3 & CXORRS J064012.8+094853 & 6 40 12.84 & 9 48 53.952 &  8.78 & 11.07 &  0.97 & 30.2 &   0.16 &  0.986E-15 & 28.83 & 100 &     \\
  4 & CXORRS J064013.5+095313 & 6 40 13.57 & 9 53 13.437 &  8.79 & 11.12 &  0.98 & 42.2 &   0.21 &  0.129E-14 & 28.95 &  99 &     \\
  6 & CXORRS J064018.3+094416 & 6 40 18.31 & 9 44 16.054 &  9.72 & 13.73 &  0.99 & 31.0 &   0.94 &  0.579E-14 & 29.60 & 100 &     \\
  8 & CXORRS J064018.5+095206 & 6 40 18.57 & 9 52  6.259 &  7.34 &  7.72 &  0.98 & 41.9 &   0.52 &  0.320E-14 & 29.35 & 100 &     \\
  9 & CXORRS J064019.3+094829 & 6 40 19.34 & 9 48 29.537 &  7.37 &  7.77 &  0.98 & 41.0 &   3.30 &  0.203E-13 & 30.15 &  65 & v,f \\
 11 & CXORRS J064020.1+094511 & 6 40 20.16 & 9 45 11.315 &  8.78 & 11.12 &  0.99 & 39.4 &   4.95 &  0.305E-13 & 30.32 &  98 &     \\
 12 & CXORRS J064020.9+094405 & 6 40 20.91 & 9 44  5.391 &  9.36 & 12.69 &  0.99 & 39.3 &   2.01 &  0.124E-13 & 29.93 &  99 &     \\
 13 & CXORRS J064020.9+094821 & 6 40 20.93 & 9 48 21.477 &  7.04 &  7.08 &  0.97 & 40.1 &   0.67 &  0.413E-14 & 29.46 & 100 &     \\
 14 & CXORRS J064020.9+095519 & 6 40 20.99 & 9 55 19.184 &  8.05 &  9.30 &  0.98 & 40.9 &   0.65 &  0.400E-14 & 29.44 & 100 &     \\
 15 & CXORRS J064021.8+095209 & 6 40 21.86 & 9 52  9.178 &  6.54 &  6.15 &  0.97 & 44.0 &   0.34 &  0.209E-14 & 29.16 & 100 &     \\
 16 & CXORRS J064022.2+095428 & 6 40 22.28 & 9 54 28.859 &  7.33 &  7.68 &  0.98 & 39.2 &  14.69 &  0.905E-13 & 30.80 &   0 & v,f \\
 18 & CXORRS J064022.6+094945 & 6 40 22.62 & 9 49 45.912 &  6.28 &  5.71 &  0.97 & 38.5 &   0.90 &  0.554E-14 & 29.58 &  99 &     \\
 19 & CXORRS J064022.9+095312 & 6 40 22.99 & 9 53 12.159 &  6.62 &  6.30 &  0.97 & 40.9 &   0.41 &  0.253E-14 & 29.24 & 100 &     \\
 20 & CXORRS J064023.3+095454 & 6 40 23.39 & 9 54 54.251 &  7.34 &  7.72 &  0.98 & 39.1 &   0.71 &  0.437E-14 & 29.48 &  99 &     \\
 22 & CXORRS J064023.7+095523 & 6 40 23.77 & 9 55 23.330 &  7.55 &  8.17 &  0.98 & 37.8 &   6.18 &  0.381E-13 & 30.42 &  96 &     \\
 23 & CXORRS J064024.8+095311 & 6 40 24.89 & 9 53 11.135 &  6.17 &  5.51 &  0.97 & 39.8 &   1.32 &  0.813E-14 & 29.75 &  99 &     \\
 24 & CXORRS J064025.5+094825 & 6 40 25.52 & 9 48 25.778 &  5.96 &  5.17 &  0.96 & 44.0 &   3.27 &  0.201E-13 & 30.14 &  99 &     \\
 25 & CXORRS J064026.9+095138 & 6 40 26.97 & 9 51 38.500 &  5.24 &  4.13 &  0.96 & 40.1 &   0.38 &  0.234E-14 & 29.21 & 100 &     \\
 26 & CXORRS J064027.3+095307 & 6 40 27.36 & 9 53  7.553 &  5.59 &  4.62 &  0.96 & 44.8 &   0.25 &  0.154E-14 & 29.03 & 100 &     \\
 27 & CXORRS J064027.6+095349 & 6 40 27.66 & 9 53 49.571 &  5.87 &  5.02 &  0.97 & 44.5 &   0.28 &  0.172E-14 & 29.08 & 100 &     \\
 28 & CXORRS J064027.8+094119 & 6 40 27.85 & 9 41 19.667 & 10.61 & 16.48 &  1.00 & 38.9 &   0.42 &  0.259E-14 & 29.25 &  99 &     \\
 29 & CXORRS J064028.8+094823 & 6 40 28.85 & 9 48 23.232 &  5.22 &  4.08 &  0.96 & 41.7 &   2.89 &  0.178E-13 & 30.09 &  99 &     \\
 30 & CXORRS J064029.0+094217 & 6 40 29.01 & 9 42 17.432 &  9.63 & 13.43 &  0.99 & 40.3 &   5.46 &  0.336E-13 & 30.37 &  50 & v   \\
 32 & CXORRS J064029.3+094407 & 6 40 29.34 & 9 44  7.148 &  8.01 &  9.20 &  0.98 & 41.8 &   0.23 &  0.142E-14 & 28.99 & 100 &     \\
 33 & CXORRS J064029.4+094736 & 6 40 29.45 & 9 47 36.838 &  5.47 &  4.43 &  0.96 & 41.4 &   0.28 &  0.172E-14 & 29.08 & 100 &     \\
 34 & CXORRS J064029.9+095010 & 6 40 29.93 & 9 50 10.049 &  4.44 &  3.15 &  0.95 & 45.4 &   1.75 &  0.108E-13 & 29.87 &  99 &     \\
 36 & CXORRS J064030.6+095014 & 6 40 30.61 & 9 50 14.085 &  4.28 &  3.00 &  0.95 & 45.5 &   0.65 &  0.400E-14 & 29.44 & 100 &     \\
 37 & CXORRS J064030.6+094610 & 6 40 30.68 & 9 46 10.398 &  6.20 &  5.56 &  0.97 & 40.4 &   3.65 &  0.225E-13 & 30.19 &  27 & v,f \\
 38 & CXORRS J064031.6+094449 & 6 40 31.62 & 9 44 49.861 &  7.11 &  7.23 &  0.98 & 39.4 &   0.35 &  0.216E-14 & 29.17 & 100 &     \\
 39 & CXORRS J064031.6+094427 & 6 40 31.68 & 9 44 27.278 &  7.44 &  7.92 &  0.98 & 39.5 &   0.89 &  0.548E-14 & 29.58 & 100 &     \\
 40 & CXORRS J064031.6+094823 & 6 40 31.70 & 9 48 23.290 &  4.64 &  3.35 &  0.95 & 40.8 &   0.25 &  0.154E-14 & 29.03 & 100 &     \\
\enddata
\tablenotetext{a}{Table available electronically.}
\tablenotetext{b}{v: variable star; f: flare-like light curve; p: possible
flare in the light curve; s: steady increase or decrease in the light curve; 
g: galaxy}
\tablenotetext{c}{Flux derived directly from best fit model of X-ray spectrum}
\end{deluxetable}

%
%
\clearpage
\begin{deluxetable}{cccccc}
\tablenum{2}
\tablewidth{0pt}
\tablecaption{Spectral Properties of Bright Sources. \label{tab_spectra}}
\tablehead{
\colhead{X-ID} &
\colhead{Optical ID} &
\colhead{$C.~R.$} &
\colhead{kT1} &
\colhead{kT2} &
\colhead{Flux\tablenotemark{a}} \\
\colhead{} &
\colhead{} &
\colhead{(counts/ks)} &
\colhead{(keV)} &
\colhead{(keV)} &
\colhead{($10^{-13}$erg/cm$^2$/s)}
}
\startdata
196 & R 3295 & 155.8 & 0.18 & 0.57 & 14.10\tablenotemark{b} \\
113 & R 2942 &  84.5 & 0.30 & 1.36 &  5.25 \\
181 & R 3245 &  54.9 & 0.76 & 3.92 &  3.45 \\
 75 & R 2752 &  42.1 & 0.80 & 5.16 &  2.48 \\
 94 & R 2840 &  36.4 & 0.40 & 2.21 &  2.34 \\
203 & R 3309 &  27.4 & 0.58 & 2.34 &  1.68 \\
142 & R 3091 &  24.1 & 1.00 & 4.46 &  1.58 \\
251 & R 3470 &  21.3 & 0.34 & 1.25 &  1.26 \\
228 & R 3390 &  18.5 & 0.64 & 2.58 &  1.04 \\
 60 & R 2633 &  18.0 & 0.30 & 1.66 &  0.79 \\
208 & R 3323 &  17.6 & 0.25 & 1.11 &  1.03 \\
230 & R 3394 &  17.2 & 0.32 & 1.62 &  0.95 \\
268 & R 3555 &  15.9 & 0.68 & 3.28 &  0.93 \\
 16 & R 2173 &  14.7 & 0.41 & 2.23 &  0.83 \\
214 & R 3342 &  13.3 & 0.32 & 2.08 &  0.76 \\
\enddata
\tablenotetext{a}{Flux determined from models of mean plasma temperatures of
0.51 keV and 2.5 keV; see text}
\tablenotetext{b}{Flux determined from best fit model; see text}
\end{deluxetable}

%
%
\clearpage
\begin{deluxetable}{cl}
\tablenum{3}
\tablewidth{0pt}
\tablecaption{Comparison of Flux Conversion Factors. \label{tab_compare}}
\tablehead{
\colhead{Conversion Factor} &
\colhead{Reference} \\
\colhead{10$^{-15}$ (ergs/cm$^2$/s)/(counts/ks)} &
\colhead{}}
\startdata
5.58      & \citet{kri01} \\
7.6-11.9  & \citet{har01} \\
8.04      & \citet{fei02} \\
6.88      & \citet{get02} \\
8.46-12.1 & \citet{dam03} \\
6.16      & this work     \\
\enddata
\end{deluxetable}

%
%
\clearpage
\begin{deluxetable}{cllrrrrrrrrr}
\tabletypesize{\scriptsize}
\tablenum{4}
\tablewidth{0pt}
\tablecaption{Optical/Infrared Sources with X-ray counterparts. \tablenotemark{a}\label{tab_optical}}
\tablehead{
\colhead{Name} &
\colhead{RA} &
\colhead{DEC} &
\colhead{X-ID} &
\colhead{U} &
\colhead{B} &
\colhead{V} &
\colhead{R} &
\colhead{${\rm I_c}$} &
\colhead{J} &
\colhead{H} &
\colhead{K} \\
\colhead{} &
\colhead{(2000)} &
\colhead{(2000)} &
\colhead{} &
\colhead{mag} &
\colhead{mag} &
\colhead{mag} &
\colhead{mag} &
\colhead{mag} &
\colhead{mag} &
\colhead{mag} &
\colhead{mag}
}
\startdata
      R 1817  & 6 40 10.00 & 9 53 41.19 &   1 &   18.47 &   18.36 &   16.93 &   16.00 &   14.94 &   13.53 &   12.94 &   12.73 \\
      R 2065  & 6 40 18.52 & 9 44 18.73 &   6 &   19.65 &   19.09 &   17.50 &   16.51 &   15.48 &   14.06 &   13.40 &   13.15 \\
      R 2066  & 6 40 18.53 & 9 52 05.73 &   8 & \nodata &   20.28 &   18.94 &   17.94 &   16.15 &   14.55 &   13.89 &   13.65 \\
      R 2093  & 6 40 19.37 & 9 48 29.83 &   9 &   18.03 &   17.38 &   15.90 &   15.00 &   13.99 &   12.59 &   11.99 &   11.89 \\
      R 2137  & 6 40 20.87 & 9 44 11.29 &  12 & \nodata &   21.94 &   20.51 & \nodata &   18.06 &   16.32 &   15.84 &   15.22 \\
      R 2163  & 6 40 21.84 & 9 52 09.11 &  15 & \nodata & \nodata &   19.92 & \nodata &   17.00 &   15.41 &   14.76 &   14.59 \\
      R 2173  & 6 40 22.28 & 9 54 28.65 &  16 &   18.20 &   17.16 &   15.62 &   14.61 &   13.56 &   12.29 &   11.56 &   11.31 \\
      R 2182  & 6 40 22.68 & 9 49 45.92 &  18 &   19.41 &   18.57 &   17.23 &   16.20 &   15.12 &   13.88 &   13.12 &   12.87 \\
      R 2190  & 6 40 23.01 & 9 53 12.08 &  19 & \nodata & \nodata &   20.50 &   18.79 &   17.16 &   15.49 &   14.78 &   14.34 \\
      R 2204  & 6 40 23.48 & 9 54 55.30 &  20 & \nodata &   20.18 &   18.54 &   17.29 &   15.44 &   14.01 &   13.35 &   12.97 \\
      R 2217  & 6 40 23.79 & 9 55 23.42 &  22 &   18.59 &   19.03 &   17.55 &   16.32 &   14.92 &   13.12 &   12.26 &   11.81 \\
      R 2251  & 6 40 24.84 & 9 53 11.45 &  23 &   19.89 &   19.23 &   17.49 & \nodata &   15.04 &   13.59 &   12.56 &   12.27 \\
      R 2271  & 6 40 25.52 & 9 48 25.88 &  24 &   17.97 &   16.83 &   15.51 &   14.66 &   13.88 &   13.24 &   12.61 &   12.45 \\
      R 2374  & 6 40 28.66 & 9 48 24.31 &  29 & \nodata & \nodata &   15.76 &   15.02 &   14.09 &   13.00 &   12.32 &   12.04 \\
      R 2383  & 6 40 29.01 & 9 42 17.18 &  30 &   17.80 &   16.79 &   15.47 &   14.66 &   13.92 &   12.94 &   12.24 &   12.05 \\
      R 2391  & 6 40 29.29 & 9 44 07.42 &  32 &   19.17 &   18.85 &   17.68 &   16.62 &   15.40 &   14.06 &   13.34 &   13.06 \\
      R 2401  & 6 40 29.46 & 9 47 36.88 &  33 &   19.79 &   20.54 &   20.15 &   17.74 &   16.77 &   14.54 &   13.89 &   13.50 \\
      R 2419  & 6 40 29.95 & 9 50 10.25 &  34 &   16.51 &   15.62 &   14.44 &   13.78 &   13.16 &   12.22 &   11.68 &   11.56 \\
      R 2442  & 6 40 30.64 & 9 50 14.38 &  36 & \nodata &   18.83 &   17.10 &   16.06 &   14.53 &   12.75 &   11.67 &   11.19 \\
      R 2443  & 6 40 30.66 & 9 46 10.59 &  37 &   10.21 &   16.63 &   15.32 &   14.61 &   13.90 &   12.86 &   12.23 &   12.06 \\
      R 2474  & 6 40 31.70 & 9 48 23.26 &  40 &   17.25 &   16.73 &   15.57 &   14.78 &   14.08 &   13.02 &   12.35 &   12.18 \\
      R 2477  & 6 40 31.73 & 9 53 30.09 &  41 & \nodata & \nodata & \nodata & \nodata &   17.31 &   14.46 &   13.11 &   12.49 \\
      R 2476  & 6 40 31.73 & 9 49 59.20 &  42 & \nodata & \nodata & \nodata & \nodata &   17.39 &   15.57 &   14.90 &   14.61 \\
      R 2507  & 6 40 32.65 & 9 49 32.90 &  44 & \nodata & \nodata &   20.02 &   18.19 &   16.37 &   14.94 &   14.28 &   14.05 \\
      R 2503  & 6 40 32.42 & 9 41 45.13 &  45 &   20.48 &   20.03 &   18.51 &   17.25 &   15.81 &   14.29 &   13.59 &   13.34 \\
      R 2511  & 6 40 32.85 & 9 51 28.97 &  46 &   18.12 &   17.06 &   15.68 &   14.78 &   13.87 &   12.68 &   12.00 &   11.79 \\
      R 2518  & 6 40 33.18 & 9 49 54.31 &  48 & \nodata & \nodata &   20.50 &   18.60 &   17.08 &   14.36 &   12.98 &   12.16 \\
      R 2550  & 6 40 33.85 & 9 48 43.64 &  50 &   19.25 &   18.62 &   17.21 &   16.11 &   14.76 &   13.45 &   12.72 &   12.35 \\
      R 2590  & 6 40 34.87 & 9 45 45.11 &  51 &   20.20 &   19.77 &   18.24 &   17.11 &   15.85 &   14.23 &   13.59 &   13.24 \\
      R 2593  & 6 40 34.94 & 9 54 06.87 &  52 & \nodata &   20.10 &   18.23 &   17.05 &   15.35 &   13.74 &   13.05 &   12.81 \\
      R 2597  & 6 40 35.23 & 9 51 56.25 &  53 &   19.00 &   17.76 &   16.20 &   15.17 &   14.01 &   12.61 &   11.83 &   11.47 \\
      R 2614  & 6 40 36.06 & 9 47 35.97 &  56 & \nodata & \nodata &   19.53 & \nodata &   16.24 &   14.29 &   13.94 &   13.51 \\
      R 2631  & 6 40 36.58 & 9 50 45.47 &  57 &   17.89 &   18.13 &   16.93 &   15.91 &   14.59 &   12.93 &   12.14 &   11.71 \\
      R 2635  & 6 40 36.69 & 9 48 22.75 &  58 &   16.94 &   16.72 &   15.61 &   14.85 &   14.06 &   13.04 &   12.33 &   12.21 \\
      R 2636  & 6 40 36.71 & 9 52 02.89 &  59 &   17.23 &   17.51 &   16.27 &   15.33 &   14.35 &   12.73 &   11.76 &   11.10 \\
      R 2633  & 6 40 36.67 & 9 47 22.53 &  60 &   13.12 &   13.20 &   12.43 & \nodata &   10.91 &    9.65 &    9.00 &    8.59 \\
\enddata
\tablenotetext{a}{Table available electronically.}
\tablenotetext{b}{Photometry from 2MASS Extended Source Catalog}
\tablenotetext{c}{Photometry contaminated by a diffraction spike}
\tablenotetext{d}{Photometry contaminated by a saturated column}
\end{deluxetable}

%
%
\clearpage
\begin{deluxetable}{rllll}
\tablenum{5}
\tabletypesize{\small}
\tablewidth{0pt}
\tablecaption{Cross identification of sources with optical/infrared 
counterparts.\tablenotemark{a} \label{tab_xids}}
\tablehead{
\colhead{X-ID} &
\colhead{R name\tablenotemark{b}} &
\colhead{Sung name\tablenotemark{c}} &
\colhead{2MASS name} &
\colhead{Other names\tablenotemark{d}} 
}
\startdata
  1 & R 1817     & \nodata    & 2MASS J06400993+0953415 &                                           \nodata  \\
  6 & R 2065     & Sung 1487  & 2MASS J06401846+0944188 &                                           \nodata  \\
  8 & R 2066     & \nodata    & 2MASS J06401847+0952060 &                                           \nodata  \\
  9 & R 2093     & Sung 68    & 2MASS J06401930+0948299 &                                           \nodata  \\
 12 & R 2137     & Sung 1526  & 2MASS J06402081+0944114 &                                           \nodata  \\
 15 & R 2163     & Sung 1539  & 2MASS J06402178+0952092 &                                           \nodata  \\
 16 & R 2173     & Sung 78    & 2MASS J06402221+0954288 &                                           \nodata  \\
 18 & R 2182     & Sung 79    & 2MASS J06402262+0949462 &                                           \nodata  \\
 19 & R 2190     & Sung 1560  & 2MASS J06402295+0953125 &                                           \nodata  \\
 20 & R 2204     & Sung 1567  & 2MASS J06402342+0954555 &                                           \nodata  \\
 22 & R 2217     & Sung 1578  & 2MASS J06402373+0955238 &                                  V594 Mon,Ogura 74 \\
 23 & R 2251     & Sung 1599  & 2MASS J06402484+0953114 &                                           \nodata  \\
 24 & R 2271     & Sung 88    & 2MASS J06402547+0948259 &                                           \nodata  \\
 29 & R 2374     & \nodata    & \nodata                 &                                           Ogura 76 \\
 30 & R 2383     & \nodata    & \nodata                 &                                          Walker 52 \\
 32 & R 2391     & Sung 1705  & 2MASS J06402924+0944075 &                                           \nodata  \\
 33 & R 2401     & Sung 1708  & 2MASS J06402941+0947369 &                                           \nodata  \\
 34 & R 2419     & Sung 104   & 2MASS J06402989+0950104 &                                   Walker 54,VSB 40 \\
 36 & R 2442     & Sung 1739  & 2MASS J06403059+0950147 &                                           \nodata  \\
 37 & R 2443     & Sung 108   & 2MASS J06403061+0946106 &                             Ogura 81,FX 15,VSB 170 \\
 40 & R 2474     & Sung 113   & 2MASS J06403164+0948233 &                                 Walker 58,V413 Mon \\
 41 & R 2477     & \nodata    & 2MASS J06403168+0953304 &                                           \nodata  \\
 42 & R 2476     & \nodata    & 2MASS J06403167+0949593 &                                           \nodata  \\
 44 & R 2507     & Sung 1773  & 2MASS J06403258+0949332 &                                           \nodata  \\
 45 & R 2503     & Sung 1767  & 2MASS J06403237+0941449 &                                           \nodata  \\
 46 & R 2511     & Sung 118   & 2MASS J06403280+0951293 &                                              FX 17 \\
 48 & R 2518     & Sung 1781  & 2MASS J06403311+0949547 &                                           \nodata  \\
 50 & R 2550     & Sung 1798  & 2MASS J06403378+0948438 &                                           \nodata  \\
 51 & R 2590     & Sung 1827  & 2MASS J06403482+0945452 &                                           \nodata  \\
 52 & R 2593     & Sung 1828  & 2MASS J06403489+0954071 &                                           \nodata  \\
 53 & R 2597     & Sung 128   & 2MASS J06403518+0951567 &                                           Ogura 85 \\
 56 & R 2614     & Sung 1841  & \nodata                 &                                           \nodata  \\
 57 & R 2631     & Sung 129   & 2MASS J06403652+0950456 &                                           \nodata  \\
 58 & R 2635     & Sung 130   & 2MASS J06403662+0948229 &                                              FX 19 \\
 59 & R 2636     & Sung 131   & 2MASS J06403665+0952032 &                                           \nodata  \\
 60 & R 2633     & Sung 132   & 2MASS J06403667+0947225 &                          Walker 66,V780 Mon,VSB 46 \\
\enddata
\tablenotetext{a}{Table available electronically.}
\tablenotetext{b}{\citet{reb02}}
\tablenotetext{c}{\citet{sun97}}
\tablenotetext{d}{Ogura \citep{ogu84}; Walker \citep{wal56};
VSB \citep{vas65}; FX \citep{fla00}; HBC \citep{her88};
Herbig \citep{her54}}
\end{deluxetable}

%
%
\clearpage
\begin{deluxetable}{llllccccc}
\tabletypesize{\scriptsize}
\rotate
\tablenum{6}
\tablewidth{0pt}
\tablecaption{Upper limits for some optical/infrared stars in the
Chandra field. \label{tab_upp_lim}}
\tablehead{
\colhead{R Name} &
\colhead{Sung Name} &
\colhead{2MASS Name} &
\colhead{Other Names} &
\colhead{Counts} &
\colhead{$C.~R.$} &
\colhead{Flux} &
\colhead{log(\lx)} \\
\colhead{} &
\colhead{} &
\colhead{} &
\colhead{} &
\colhead{(counts)} &
\colhead{(counts/ks)} &
\colhead{($10^{-13}$erg/cm$^2$/s)} &
\colhead{log(erg/s)}
}
\startdata
R 2772 & Sung 153 & 2MASS 06404102+0947577 & \nodata 
  & $<$ 4.11 & $<$ 0.0955 & $<$0.0056 & $<$ 28.61 \\

R 2898 & Sung 173 & 2MASS 06404464+0948021 & Walker 90, V590 Mon, HBC 219, VSB 62
  & $<$ 3.00 & $<$ 0.0703 & $<$0.0041 & $<$ 28.48 \\

R 2982 & Sung 2066 & 2MASS 06404729+0947274 & \nodata  
  & $<$12.63 & $<$ 0.3940 & $<$0.0231 & $<$ 29.22 \\

R 2491 & Sung 1760 & 2MASS 06403200+094935 & \nodata
  & $<$ 3.00 & $<$ 0.1794 & $<$0.0105 & $<$ 28.88 \\

\nodata & \nodata & \nodata & V590 Mon, HBC 219, Herbig 25
  & $<$ 4.11 & $<$ 0.0962 & $<$0.0056 & $<$ 28.61 \\

R 2792 & Sung 159 & 2MASS 06404156+0955174 & VSB 56
  & $<$ 3.00 & $<$ 0.1984 & $<$0.0116 & $<$ 28.93 \\

\nodata & \nodata & \nodata & LN Mon, HBC 215, Herbig 21
  & $<$ 4.82 & $<$ 0.1193 & $<$0.0070 & $<$ 28.71 \\

R 3167 & Sung 2178 & 2MASS 06405413+0948434 & \nodata
  & $<$ 4.11 & $<$ 0.0944 & $<$0.0055 & $<$ 28.60 \\

R 3216 & Sung 2206 & 2MASS 06405573+0946456 & \nodata
  & $<$ 3.82 & $<$ 0.0903 & $<$0.0053 & $<$ 28.58 \\

\nodata & \nodata & \nodata & LR Mon, HBC 220, Herbig 27
  & $<$ 4.11 & $<$ 0.1816 & $<$0.0106 & $<$ 28.89 \\

R 2972 & Sung 2056 & 2MASS 06404694+0955036 & \nodata
  & $<$ 7.24 & $<$ 0.1697 & $<$0.0099 & $<$ 28.86 \\

R 3022 & Sung 195 & MASS 06404888+0951444 & Walker 100, HD261841, VSB 72
  & $<$ 7.40 & $<$ 0.1677 & $<$0.0098 & $<$ 28.85 \\

R 3041 & Sung 196 & 2MASS 06404953+0953230 & Walker 104, VSB 74
  & $<$ 6.78 & $<$ 0.1561 & $<$0.0092 & $<$ 28.82 \\

R 3073 & Sung 2125 & 2MASS 06405033+0954158 & \nodata
  & $<$ 4.11 & $<$ 0.0958 & $<$0.0056 & $<$ 28.61 \\

R 3110 & Sung 209 & 2MASS 06405155+0951494 & Walker 109, HD261878, VSB 79
  & $<$ 5.82 & $<$ 0.1328 & $<$0.0078 & $<$ 28.75 \\

\nodata & \nodata & \nodata & FX 67, VSB 245
  & $<$ 3.94 & $<$ 0.0926 & $<$0.0054 & $<$ 28.60 \\

\enddata
\end{deluxetable}

\end{document}